\documentclass[11pt,twoside]{article}
\usepackage{cozumel2005}
\usepackage{epsf}
\usepackage{psfig}
\usepackage{lscape}
\begin{document}
\title{Photometric entropy of stellar populations and related diagnostic tools}    
\author{Alberto Buzzoni}   
\affil{INAF - Osservatorio Astronomico di Bologna, Via Ranzani 1 - 40127 Bologna (Italy)}    

\begin{abstract} 
We discuss, from a statistical point of view, some leading issues that deal with the study of
stellar populations in fully or partially unresolved aggregates, like globular clusters and 
distant galaxies.
A confident assessment of the effective number and luminosity of stellar contributors 
can provide, in this regard, a very useful interpretative tool to properly assess the observational 
bias coming from crowding conditions or surface brightness fluctuations.
These arguments have led us to introduce a new concept of ``photometric entropy''
of a stellar population, whose impact on different astrophysical aspects of cluster diagnostic 
has been reviewed here.
\end{abstract}

\section{Introduction}
Galaxy surface brightness fluctuations and crowding effects in nearby star clusters 
are two related and well recognized features one has to deal with when observing partially
or fully unresolved stellar systems.
The latter effect turns out to be a severe problem, for instance, when probing the
innermost regions of  globular clusters in our own galaxy, and more generally in all those 
situations when the stellar plot consists in fact of blended point sources, at different spatial scales. 

On the other hand, to a more detailed analysis, even the smooth surface brightness of distant  
galaxies actually is found to display some intrinsic ``clumpiness'',  and this special property can be 
usefully exploited to derive information on their fully blended composing stellar populations, 
as first shown by \citet{ts88} and \citet{t91}.

In this contribution, we would like to further extend the analysis of these two important issues, 
that deeply relates to overall characteristics of stellar aggregates and the way the latter 
are sampled by the observations. 
In particular, we will show that both effects are in fact a consequence of the same property
of {\it finiteness} and {\it discreteness} of the composing stellar populations, and this will lead 
to a unified and more general definition of ``photometric entropy'' of a stellar population, a relevant 
concept whose impact on different astrophysical aspects of cluster diagnostic we want to review here.

\begin{figure}[!t]
\centerline{
\psfig{file=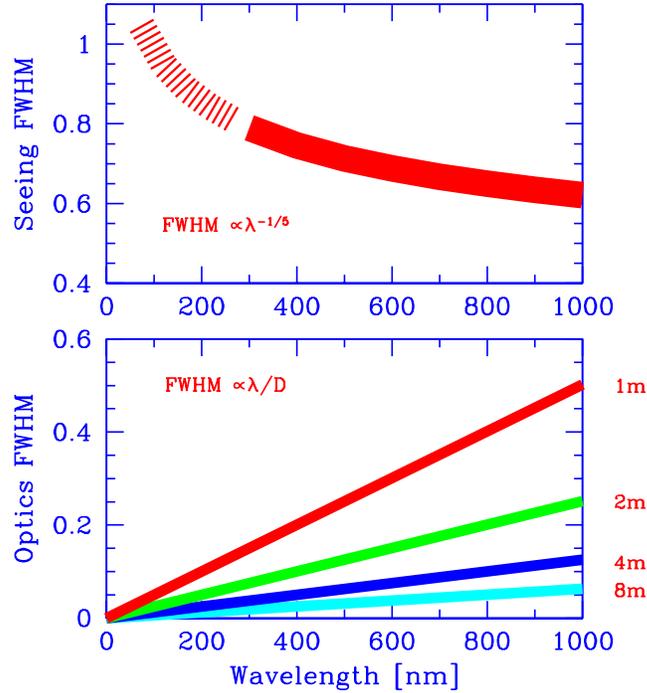,width=0.7\hsize,clip=}
}
\caption{{\it Upper panel} - The expected trend of seeing variation with varying wavelength,
after \citet{w99}. From empirical estimates and theoretical arguments, the FWHM of the
seeing disk is found to scale approximately as FWHM$_{\rm seeing}\propto \lambda^{-1/5}$.
{\it Lower panel} - Telescope resolving power with varying the aperture
diameter, $D$, (right labels, in meters). As predicted by the classical Airy diffraction pattern,
FWHM$_{\rm Airy} \propto \lambda/D$. Contrary to seeing, optics performance degrades when
moving to long-wavelength photometric bands.}
\label{fig1}
\end{figure}

\section{Observational constraints}

Our attempt to single out any ``clean'' information of a cluster stellar population crucially relies 
on how confidently we approach the ideal observing conditions, that is those allowing us to fully 
resolve any star member in the system. Apart from the obvious constraint due to the 
physical distance of our target and its apparent size,
there are at least three relevant external mechanisms that play a role to bias our results 
when carrying out imagery (or spectroscopy) of a stellar aggregate.
\begin{itemize}
\item {\underline{\bf Photon noise}} directly relates to the signal-to-noise ratio ($S/N$)
of our observations, the latter being constrained both by telescope size and,
in case of ground-based observations, by sky brightness. In particular, we know that, for faint 
sources, $S/N$  mainly scales with the telescope aperture ($D$), the exposure time ($t_{\rm exp}$), 
and the sky surface brightness (ssb), in magnitudes per square arcsec, such as 
\begin{equation}
S/N \propto (1/D)\times (1/\sqrt{t_{\rm exp}}) \times 10^{0.2\,{\rm ssb}}.
\end{equation}

\item{\underline{\bf Telescope resolving power}} is also an issue.
It can be quantified by the full width at half maximum (FWHM) of the instrumental diffraction pattern
(the ``Airy disk''), that is a measure of the {\it minimum} physical ``spot'' of 
a star on our imagery detector (or spectrograph slit), if we could observe outside the atmosphere. 
The Airy disk depends itself on $D$, but scales with the observing wavelength, as well, so that 
\begin{equation}
{\rm FWHM}_{\rm Airy} \simeq {\lambda \over D}.
\end{equation} 

\item Finally, {\underline{\bf seeing conditions}}, in case of ground-based observations,
drastically depend on sky clarity and, opposite to the telescope resolving power, improve in general at longer wavelength, 
mainly due to a better spatial coherence of the atmosphere convection layers
with increasing $\lambda$.  Theoretical arguments and empirical estimates \citep[][]{w99}
indicate that FWHM of the seeing disk scales as 
\begin{equation}
{\rm FWHM}_{\rm seeing} \propto \lambda^{-1/5}.
\end{equation}
\end{itemize}

While, to some extent, we could improve $S/N$ of our observations (and ``see more clearly'') 
by arbitrarily increasing exposure time 
(although with some non-negligible technical limits due to detector saturation, electronic noise etc.), 
the other two effects are much more difficult (and expensive) to be optimized, as one would need either a bigger
telescope and/or space-based observations to fully remove any atmosphere contamination.
The combined action of telescope resolving power and seeing provides therefore the observational 
kernel that constrains our ability to master the crowding conditions and fully resolve the stellar 
population (see Fig.~\ref{fig1}). 

For example, for the typical case of a $\sim50$~pc globular cluster around the Andromeda galaxy 
(about 0.7~Mpc away) a  mean angular separation of the order of $0.04\arcsec$ should be expected 
among the $\sim 10^5$ star members of the cluster. Such an equivalent resolving 
power would certainly not be achieved by means of any seeing-limited conventional telescope on 
the ground, rather demanding at least a 2.5~m orbiting telescope (as it is actually the case for 
the Hubble Space Telescope, indeed!).

\begin{figure}
\centerline{
\psfig{file=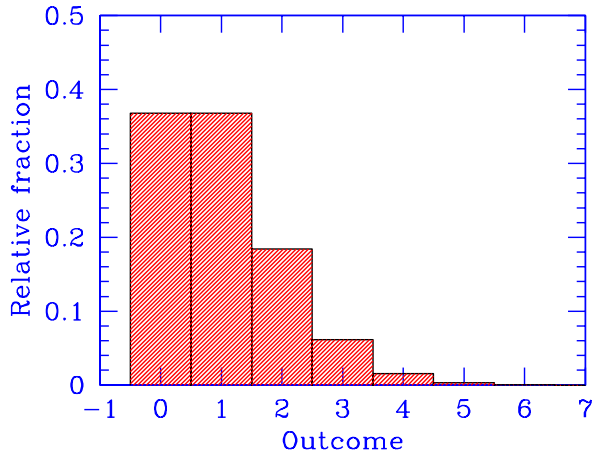,width=0.64\hsize,clip=}
}
\caption{The Poissonian distribution of the $N_{\rm tot}$ star cells in our 
``fair statistical representation'' of a stellar population. In average, each cell contains
one star, according to the Poisson formula, $p(x) = {{e^{-1}}/x!}$. As a consequence, after repeated 
trials, the total number of stars, $N_{\rm tot}$, is expected to fluctuate by 
$\sigma(N_{\rm tot}) = \sqrt{N_{\rm tot}}$.}
\label{fig2}
\end{figure}

\section{Effective numbers of luminous contributors and photometric entropy}

Although within the limits of the instrumental performance and seeing influence
(both being, however, quantitatively assessed and, at least partially, recovered),
one can still take advantage of the overall statistical properties of a stellar sample
and gain valuable information, in any case, of its unresolved component.

In this regard, photometric entropy adds an important tool to our analysis. As we will see
in a moment, its definition has much to do with the concept of
``effective number'' of stellar contributors to the integrated luminosity of a stellar
population. This quantity has been first assessed in a series of previous papers
\citep{b89,b93}, providing the reference framework for further theoretical investigations
\citep[][]{c00,c01,c02,cl04}. We will just sketch here some
of the leading issues of the theory and the main relationships among the relevant statistical 
quantities involved in our analysis.

First of all, it is important to define operationally what we call a ``fair statistical 
representation'' of a stellar population. Ideally, we could assume to have a number $N_{\rm tot}$
of cells, each to host, {\it in average}, one star to be supplied to the system through
a stochastic process of Poissonian nature. After completion of $N_{\rm tot}$ iterations, 
each cell will display a star distribution ($n_j$) like in Fig.~\ref{fig2}, and the whole system 
will contain, {\it in average}, $\sum_{j=1}^{N_{\rm tot}} n_j = N_{\rm tot}$ stars. 

Due to the Poissonian distribution of $n_j$, however, in repeated statistical realizations
of the population, the total number of stars will fluctuate by $\sigma(N_{\rm tot}) = 
\sqrt{N_{\rm tot}}$.

If we assume, to a first analysis, that all stars have the same luminosity $\ell$,
then the expected relative fluctuation of the global luminosity ($L_{\rm tot}$) of the sample is:
\begin{equation}
{{\sigma(L_{\rm tot})}\over {L_{\rm tot}}} = {{\sqrt{\sum_j n_j^2 \ell^2}}\over {\sum_j n_j \ell}} =
{{\ell \sqrt{N_{\rm tot}}} \over {\ell N_{\rm tot}}} = 1/\sqrt{N_{\rm tot}}.
\label{eq:sigma}
\end{equation}

More likely, if $\ell_j$ is not a constant, and distributes according to a given
luminosity function, then we could still retain eq.~(\ref{eq:sigma}), just
replacing $N_{\rm tot}$ in the r.h. side of the formula with a more general quantity $N_{\rm eff}$.
The value of $N_{\rm eff}$ can be regarded as an ``effective'' number of stellar
contributors in the population; it could be demonstrated that, always, 
\begin{equation}
N_{\rm eff} \le N_{\rm tot}.
\end{equation}
Quite importantly, as the luminosity distribution of stars changes with wavelength,
one has to expect that 
\begin{equation}
N_{\rm eff} = N_{\rm eff}(\lambda).
\end{equation}

\begin{figure}
\centerline{
\psfig{file=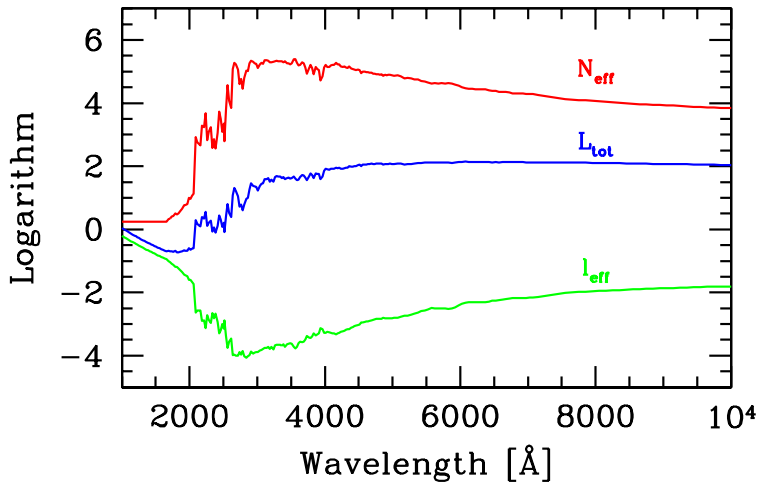,width=0.8\hsize,clip=}
}
\caption{Logarithm of effective luminosity ($\ell_{\rm eff}$) and star number
($N_{\rm eff}$) for a 15~Gyr SSP of solar metallicity and Salpeter IMF, from \citet{b93}.
Both quantities relates to the global luminosity ($L_{\rm tot}$) of the stellar population
through eq.~(\ref{eq:nll}). The model has been rescaled to a total SSP bolometric luminosity
of $10^7$\,L$_\odot$.}
\label{fig3}
\end{figure}

On a similar argument, the {\it relative variance} of the luminosity
($L_{\rm tot}$) for the whole stellar population simply results:
\begin{equation}
{{\sigma^2(L_{\rm tot})}\over {L_{\rm tot}}} = {{\sum_j n_j \ell_j^2}\over {\sum_j n_j \ell_j}} =
\ell_{\rm eff}.
\label{eq:variance}
\end{equation}
In the equation, $\ell_{\rm eff}$ is the ``effective luminosity'', that is basically a ``mean'' 
representative luminosity of the composing stars, using $w_j = n_j\ell_j/L_{\rm tot}$ as a 
normalized weighting factor.

It is interesting to note that eq.~(\ref{eq:variance}) is the key relation for the \citet{ts88}
theory of galaxy surface brightness fluctuations. 
Actually, one major issue of the Tonry \& Schneider method is that, relying on the observed
galaxy flux $f_{\rm gal}$, one can supply the empirical quantity
\begin{equation}
{{\sigma^2(f_{\rm gal})}\over {f_{\rm gal}}} = {\ell_{\rm eff}\over {4\,\pi\,d^2}}
\end{equation}
(being $d$ the galaxy distance), to be matched with the corresponding theoretical predictions 
for $\ell_{\rm eff}$ from population synthesis models, according to the distinctive properties 
of the (unresolved) galaxy stellar population.
As a result, the method proves to be, in principle, a powerful distance indicator, leading to
a direct measure of the galaxy distance modulus:
\begin{equation}
(m-M) = 5\,\log d -5 = 5\,\log \sqrt{\left({{\ell_{\rm eff}}\over {4\,\pi}}\right)
\left({{f_{\rm gal}\over \sigma^2(f_{\rm gal})}}\right)} -5.
\end{equation}

By matching eqs.~(\ref{eq:sigma}) and (\ref{eq:variance}), a relevant
physical constraint should hold (see Fig.~\ref{fig3}) as:
\begin{equation}
N_{\rm eff} \times \ell_{\rm eff} = L_{\rm tot}.
\label{eq:nll}
\end{equation}

\begin{figure}
\centerline{
\psfig{file=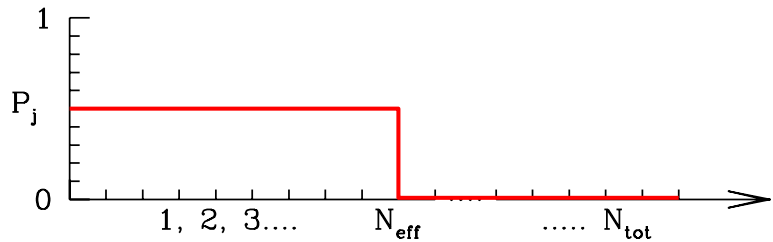,width=0.8\hsize,clip=}
}
\caption{A schematic representation of $N_{\rm eff}$, defined in our statistical experiment as the 
{\it maximum} number of  ``switched-on'' cells, each containing one star of fixed luminosity
$\ell_{\rm eff}$, such as to provide the total observed luminosity of the population, $L_{\rm tot}$,
at a given photometric band.}
\label{fig4}
\end{figure}

Accordingly, from a statistical point of view, $N_{\rm eff}$ represents therefore
{\it the maximum number of bright stars, of constant luminosity $\ell_{\rm eff}$, 
allowed in a population of $N_{\rm tot}$ members to provide the total luminosity $L_{\rm tot}$} 
(see Fig.~\ref{fig4}). Ideally, this statistical definition assumes a ``two-state'' condition for
sample stars, with $N_{\rm eff}$ ``switched-on'' objects of $\ell_{\rm eff}$ individual luminosity
and the remaining $(N_{\rm tot} - N_{\rm eff})$ ``switched-off'' stars with $\ell = 0$. 
Alternatively, $N_{\rm eff}$ can also be regarded as the {\it maximum} number of states 
(i.e.\ ``switched-on'' cells in our previous example) available to the system, and this 
straightforwardly leads to a more general extension of the ``entropy'' ($S$) concept, that 
in our framework can now be defined as 
\begin{equation}
S = \log (N_{\rm eff}/N_{\rm tot}).
\end{equation}
or, in its usual thermodynamical notation,
\begin{equation}
\Delta S = \Delta \log N_{\rm eff}
\end{equation}
as we would better like to single out its variation within a stellar population rather than 
its absolute value.

According to our normalization, as $0 < N_{\rm eff}(\lambda)/N_{\rm tot} \le 1$,
$S$ is always a negative quantity, depending on wavelength and tipping at zero, in case
of a ``maximum-entropy'' stellar system, where all stars contribute with the same luminosity
at a given photometric band.

\begin{figure}
\centerline{
\psfig{file=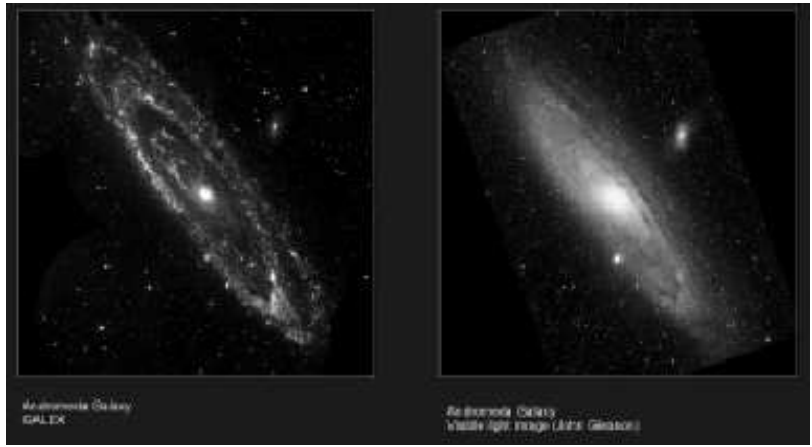,width=0.8\hsize,clip=}
}
\caption{The centermost $3\deg \times 3\deg$ region of the Andromeda galaxy (M\,31), as imaged 
by the ultraviolet space telescope GALEX \citep[][]{th05} between 1500 and 3000~\AA\ {\it (left panel)}, and seen at optical 
wavelength {\it (right panel)}. The image is courtesy of the GALEX scientific team, and made 
available at the Project Web site {\sf http://www.galex.caltech.edu/}. It is evident the
increased ``transparency'' of the galaxy at ultraviolet wavelength, as a consequence of
a drastically lower number of effective stellar contributors. See text for a discussion.}
\label{fig5}
\end{figure}

\section{Crowding effects and apparent opacity of stellar systems}

As shown in Fig.~\ref{fig3}, one striking feature of the $N_{\rm eff}$ function (and its derived 
entropy quantity) is its wide change with wavelength. In particular, one sees that
for a 15 Gyr simple stellar population (SSP) model, from \citet{b93}, the entropy peaks 
around blue/visual wavebands ($UBV$), while the number of effective
contributors dramatically drops by nearly five orders of magnitude when moving to short wavelength,
below 3000~\AA. A similar (although much milder) trend is also evident at infrared wavelength, where
$N_{\rm eff}$ smoothly decreases by about one dex, compared to the optical range. 

Scaled to a typical $10^{11}$\,L$_\odot$ galaxy, this means 
that, by looking at mid- ultraviolet wavelength, we have to expect a thin plot
of some $10^4$ UV-bright stars tracing the galaxy body, compared to
a quite smooth surface brightness distribution at visual wavelength, provided by 
about $10^{10}$ effective stellar contributors. 
The Andromeda galaxy, as imaged in the visible light and by the ultraviolet space telescope 
GALEX \citep[][]{th05}, is a good example in this sense (see Fig.~\ref{fig5}).

The exact trend of $N_{\rm eff}$ (and $S$) vs.\ wavelength is a natural output of theoretical codes 
for population synthesis, and it can easily be computed for a wide range of the 
distinctive evolutionary parameters for a stellar population. Detailed values for SSP grids
of models can be found in \citet[][and Web updates at {\sf http://www.bo.astro.it/~eps/home.html}]{b93} 
and \citet{c02}.

The knowledge of the entropy level is of special importance in order to quantitatively assess
the expected crowding conditions or rather the apparent ``optical depth'', when observing a 
distant stellar system at different photometric bands. The latter will depend in fact on the 
effective number of stellar contributors convolved with the instrumental kernel (telescope 
diffraction pattern plus seeing PSF) that degrades our resolution leading to a smooth
nearly ``solid'' surface brightness profile.

\begin{figure}[!t]
\centerline{
\psfig{file=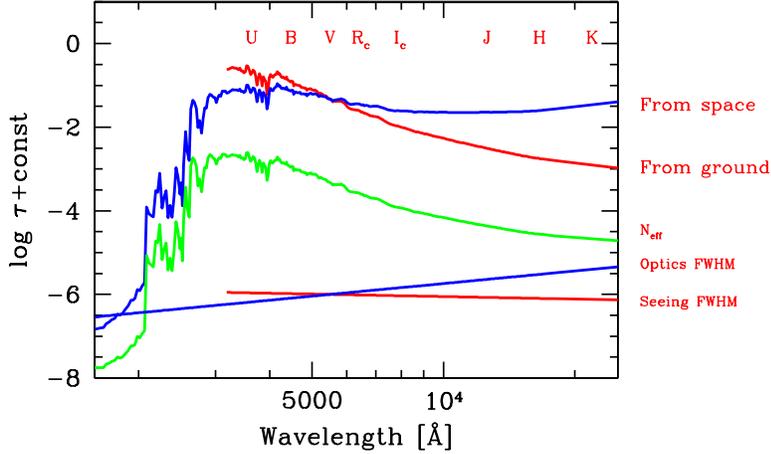,width=0.8\hsize,clip=}
}
\caption{The apparent system opacity ($\tau$) in case of a SSP as 
in Fig.~\ref{fig3}, assuming to observe the stellar sample from ground and from space (upper curves).
Crowding conditions are mainly modulated by seeing, in case of ground-based observations 
with conventional telescopes, and by the telescope diffraction limit in case of space imaging.
Note that calculations for ground-based observations have been stopped shortward of 
3000~\AA\ given the atmosphere opacity. In addition, the two curves have been offset by an
arbitrary quantity in the logarithm domain to match at the $V$ band.
Due to an opposite trend of seeing and optics FWHM (see the two curves at the bottom, as
labeled), the convolution with the apparent surface density of stars ($N_{\rm eff}$ curve)
leads to different performance outputs. In particular, ground-based observations
sensibly improve when moving at longer wavelength, thanks to a more favorable seeing
condition, while crowding conditions are expected to maintain roughly constant at the
different photometric bands in case of space observations. The Johnson/Cousins photometric
band system is reported at the top of the figure, as a reference.}
\label{fig6}
\end{figure}

In case of ground-based (i.e.\ seeing limited) imaging of a star cluster, for instance, the 
apparent optical depth ($\tau$) is expected to scale as
\begin{equation}
\log \tau_{\rm ground}  = S+\log{{\pi\,{\rm FWHM}_{\rm seeing}^2}\over {{\rm Cluster~apparent~area}}} +{\rm const}.
\end{equation}
If, on the contrary, we are observing  from space (that is fully exploting the telescope 
diffraction limit), then target opacity will scale as
\begin{equation}
\log \tau_{\rm space} = S+\log{{\pi\,{\rm FWHM}_{\rm Airy}^2}\over {{\rm Cluster~apparent~area}}} +{\rm const}.
\end{equation}

In this framework, the difference between a ``crowded field'' and a ``surface brightness 
fluctuation'' is just a matter of apparent ``opacity'' of the stellar system, eventually 
depending whether $\tau$ is respectively $\ll 1$  or $\gg 1$.

According to the trend envisaged in Fig.~\ref{fig1}, the expected performance
of our observations will therefore change with $\lambda$ as in Fig.~\ref{fig6}.
In general, we see that ground-based observations are more sensitive to the
observing photometric band as stars will more easily be 
picked up at longer wavelength, taking advantage of seeing improvement; on the contrary, 
our output will not be so critically constrained when observing from space,
as the lower value of $N_{\rm eff}$ at infrared wavelength barely compensates the poorer telescope 
resolving power (compared to the visible wavelength range), thus maintaining the crowding 
conditions roughly unchanged along the different photometric bands.

\begin{figure}[!t]
\centerline{
\psfig{file=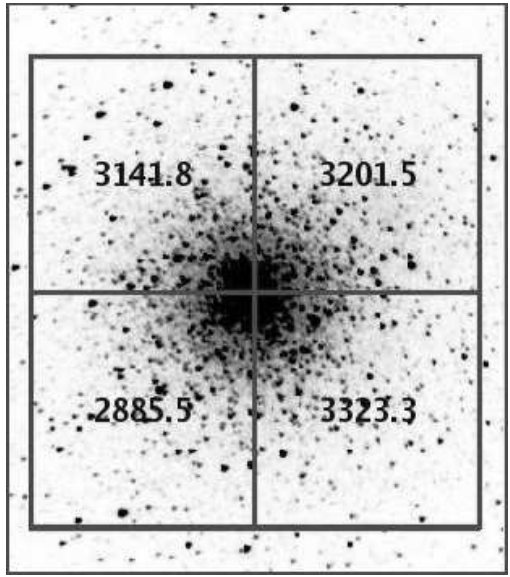,width=0.5\hsize,clip=}
}
\caption{An illustrative example of surface brightness fluctuations in the globular clsuter M\,53
(alias NGC\,5024). The $R_c$-band frame has been taken at the 2.12m telescope of Cananea (Mexico)
with a FOSC camera. Field is roughly $6.5\arcmin\times6.5\arcmin$ sampled with a $1\arcsec$  pixel size of
on the sky. For our experiment, the cluster has been divided in four quadrants reporting
the sky-subtracted integrated CCD counts for each partition, as labeled on the plot. See text for
discussion.}
\label{fig7}
\end{figure}

\section{Surface brightness fluctuations}

The previous arguments in our discussion make easier now a different statistical 
approach to the problem of surface brightness fluctuations in  stellar systems.
In particular, relying on our definition of ``fair statistical realization'' of
a stellar sample (see Sec.\ 3), one could take advantage of cluster and galaxy symmetry
for a different and more general statement of the \citet{ts88} method. 

For example, in case of a globular cluster, like the $R_c$-band CCD picture of M\,53 
shown in Fig.~\ref{fig7}, one could assume that any quadrant of the system (by centering on 
the photometric barycenter of the cluster) collects a ``statistically fair sample'' of the 
whole cluster population, in force of the claimed central symmetry of the system. 
So, while, {\it in average}, each quadrant will display  a mean luminosity $L'= L_{\rm tot}/4$, 
a {\it scatter} has to be expected for this quantity, according to l.h.\ side of 
eq.~(\ref{eq:variance}), simply measured as the variance of the four count determinations.
In the specific case of Fig.~\ref{fig7} we have
\begin{equation}
\left\{
\begin{array}{ll}
L' = 3138 \quad {\rm counts} \\
\sigma^2(L') = 184^2 \quad {\rm counts}.
\end{array}
\right.
\end{equation}
The apparent ($R_c$-band) effective  magnitude ($m_{\rm eff}$) of the M\,53 stellar population 
becomes therefore
\begin{equation}
m_{\rm eff} = -2.5\,\log {\ell_{\rm eff}\over {4\,\pi\,d^2}} = -2.5\, \log 
\left({\sigma^2(L')\over L'}\,{1\over {4\, L'}}\right) + M_R(\rm M\,53),
\end{equation}
being $L_{\rm tot} = 4\,L'$ the apparent luminosity of the cluster (i.e.\ across the four quadrants, in CCD counts),
to be calibrated in magnitude scale through the known apparent $R_c$-band magnitude of M\,53 
[$M_R(\rm M\,53) = 7.11$, from \citet{h96}, dereddened assuming a color excess $E(B-V) = 0.02$ and
$A(R_c) = 2.6 E(B-V)$ from \citet{mm00}].
Replacing the relevant quantities, for our cluster we have $m_{\rm eff} = 14.78$~mag.

For the M\,53 stellar population, the \citet{b89,b93} SSP models predict and absolute
($R_c$-band) effective magnitude $M_{\rm eff} = -1.58$~mag, assuming an age of $\sim 15$~Gyr
for the cluster, a metallicity [Fe/H]~$= -1.88 \pm 0.2$~dex \citep{sp04} and 
a blue horizontal branch (HB) morphology. This eventually leads to an estimated distance modulus
for M\,53 of
\begin{equation}
m_{\rm eff} - M_{\rm eff} = 14.78 + 1.58 = +16.36~{\rm mag}
\end{equation}
in quite good agreement with the standard value of $(m-M) = 16.25$~mag from
the \citet{h96} globular cluster catalog; this implies a $\sim 5$\% relative uncertainty 
in the derived distance of the cluster with our method.

\subsection{Color fluctuations and stellar sampling}

Previous application of cluster ``fair sampling'' basically relies on the claimed
symmetry of the system. In this regard, our adopted 4-quadrant partition is fully
arbitrary, as any other geometrical combination, like slices of fixed angular aperture,
or other fixed-size apertures symmetrically located around the center  could 
work equally well in our method.

As an interesting variant on this line, that may more suitably apply to fully unresolved galaxies,
one could also consider to trace the luminosity variance along a given isophote of
the surface brightness profile. This could easily be done, for instance, in case of ellipticals 
or face-on spirals.

One important consequence of the finite luminosity sampled per pixel resolution element,
along a given galaxy isophote, is that some {\it intrinsic} variance of the apparent color 
has to be expected, as a consequence of a different photometric entropy with varying the observing
band. The amplitude of this color scatter depends on the amount of bolometric $L_\odot$ 
sampled per pixel and the relative variation of $N_{\rm eff}$ at the different photometric bands.

\begin{figure}
\centerline{
\psfig{file=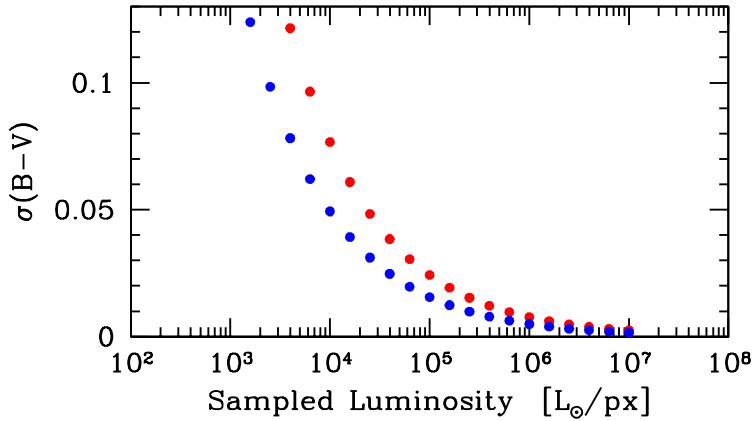,width=0.8\hsize,clip=}
}
\caption{The expected statistical scatter in the galaxy B-V color versus pixel luminosity sampling.
The scatter amount ranges according to eq.~(\ref{eq:color}). High-$S/N$ (i.e.\ $S/N \gg 100$)
high-resolution galaxy imagery could in principle detect the effect.}
\label{fig8}
\end{figure}

An example is done in Fig.~\ref{fig8}, where we computed the statistical variation of 
the integrated B-V color versus sampled luminosity. The B-V variance (in magnitude scale) 
derives from 
\begin{equation}
\sigma^2(B-V) = \sigma^2(B) + \sigma^2(V) -2\rho\,\sigma(B,V)
\label{eq:sbv}
\end{equation}
where $\sigma^2(B) = 1/N_{\rm eff}^B$, $\sigma^2(V) = 1/N_{\rm eff}^V$,
and the covariance term $-2\rho\,\sigma(B,V)$ ranges between zero (if the $B$ and $V$ luminosity 
contributors can be assumed to be totally independent, that is for a
correlation coefficient $\rho = 0$ in eq.~\ref{eq:sbv}), and  $-2\sigma(B)\sigma(V)$ if we assume
a perfect positive correlation between the two photometric bands (i.e.\ assuming $\rho = 1$).
As shown in Fig.~\ref{fig8}, the full variation range for the statistical fluctuation of the 
integrated color can eventually be written as 
\begin{equation}
\mid\sigma(B) - \sigma(V)\mid \,\le \sigma(B-V) \le \sqrt{\sigma^2(B)+\sigma^2(V)}
\label{eq:color}
\end{equation}

Quite interestingly, one sees from the figure that a detectable color scatter could be
measured along a galaxy surface isophote providing to collect high-($S/N$) (i.e.\ $S/N \gg 100$)
high-resolution imagery.
For example, in case of a typical $10^{11}$\,L$_\odot$ elliptical galaxy at the Virgo distance 
($d \simeq 15$~Mpc), one expects to sample roughly $10^5$~L$_\odot$/px with a CCD of $\sim 0.2\arcsec$ 
pixel size. This leads to a fully measurable $\sigma(B-V)$ scatter of some 0.01~mag (see
Fig.~\ref{fig8}). The possible cosmological relevance of this test is obvious, as from a 
measure of a color scatter in a distant galaxy we can derive an absolute estimate of the 
sampled luminosity, to be compared with the observed surface brightness and derive therefrom
the luminosity distance.

\begin{figure}[!t]
\centerline{
\psfig{file=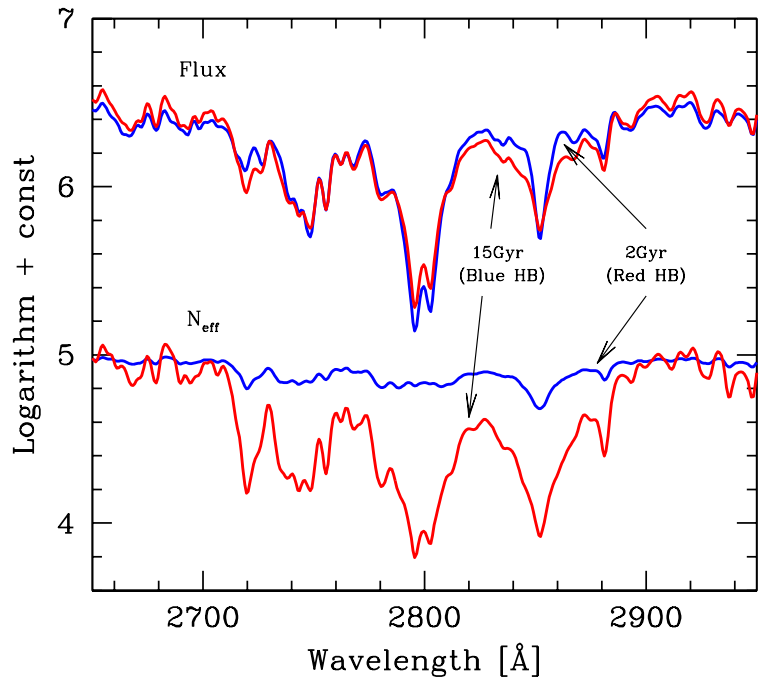,width=0.8\hsize,clip=}
}
\caption{Theoretical spectral energy distribution {\it (upper plots)} and effective stellar 
contributors ($N_{\rm eff}$) {\it (lower plots}) for two SSPs of solar metallicity and Salpeter 
IMF according to Buzzoni (1989, 1993) population synthesis models.
The models refer to a 2~Gyr SSP with red HB morphology, and a 15~Gyr population
with blue HB, as labeled. The new {\sc Uvblue} theoretical library of spectra \citep[][]{rcbb05}
has been used to reproduce the spectral region around the striking Mg{\sc ii} and Mg{\sc i} features,
about 2800~\AA. In spite of a substantially indistiguishable difference between the spectra, the two
SSPs display a large difference in terms of effective contributors and photometric entropy,
with a much higher statistical scatter in the spectral features expected for the older SSP.}
\label{fig9}
\end{figure}

\section{Diagnostic tools for high-resolution spectroscopy}

Following a substantially similar argument as for the color statistical scatter of previous
section, one could also further expand the approach including spectroscopy.
Again, the change of $S$ and $N_{\rm eff}$ along wavelength is the key issue that leads us to
expect some variance among the spectral features, even at close wavelength difference.

A confident assessment of this effect could be extremely useful as an additional interpretative 
tool to disentagle the well recognized degeneracy among the evolutionary parameters of
stellar populations (see, e.g.\ the long-lasting question of the ``age-metallicity'' dilemma
as discussed, for instance, by \citealp{rb86} and \citealp{b95}).

Figure~\ref{fig9} is an illuminating example of this kind of problem. The two plots in the figure
display the ultraviolet spectral energy distribution and effective number of stellar contributors for
two SSPs of different age and HB morphology. The models have been obtained by matching
the \citet{b89} synthesis code with the new {\sc Uvblue} theoretical spectral library of
\citet{rcbb05}. 

In particular, a young (2~Gyr) population with red HB morphology is compared with an old 
(15~gyr) one, with a blue HB. One sees that, in spite of the striking age difference, the two 
spectra are nearly identical as, in both cases, we have a dominating component of warm stars 
($T_{\rm eff} \sim 10\,000$~K) that emit in the ultraviolet range. However, while in the 2~Gyr 
SSP the UV-luminosity is provided by a large number of main sequence turn-off stars, in the 15~Gyr 
case we have a prevailing contribution from a few bright stars in the blue tail of the HB.

This has a direct impact on the value of $N_{\rm eff}$, with a much larger scatter along
wavelength for the oldest SSP. As a consequence, in the latter case we should likely expect a larger
variance (about a factor of two) in the measure of any narrow-band spectrophotometric index 
along the galaxy isophotes [including both the popular Lick indices of \citet{w94}, in the 
optical range or the \citet{f90} ultraviolet indices], in a way very similar to what we 
have shown for the B-V color.

\acknowledgements 
It is a pleasure to thank the organizers, David Valls-Gabaud and Miguel Chavez, for their 
kind invitation to attend this exciting workshop.


\begin{thebibliography}{}
\bibitem[Buzzoni(1989)]{b89} Buzzoni, A. 1989, \apjs, 71, 817
\bibitem[Buzzoni(1993)]{b93} Buzzoni, A. 1993, \aap, 275, 433
\bibitem[Buzzoni(1995)]{b95} Buzzoni, A. 1995, \apjs, 98, 69
\bibitem[Cervi\~no \& Luridiana(2004)]{cl04} Cervi\~no, M. \& Luridiana, V. 2004, 413, 145
\bibitem[Cervi\~no, Luridiana \& Castander(2000)]{c00} Cervi\~no, M., Luridiana, V. \& Castander, F. J. 2000, \aap, 360, 5
\bibitem[Cervi\~no et al.(2001)]{c01} Cervi\~no, M., G\'omez-Flechoso, M. A., Castander, F. J., Schaerer, D., Moll\'a, M., Kn\"odlseder, J. \& Luridiana, V. 2001, \aap, 376, 422
\bibitem[Cervi\~no et al.(2002)]{c02} Cervi\~no, M., Valls-Gabaud, D., Luridiana, V. \& Mas-Hesse, J. M. 2002, \aap, 381, 51
\bibitem[Fanelli et al.(1990)]{f90} Fanelli, M.N., O'Connell, R.W., Burstein, D. \& Wu, C.C. 1990, 364, 272
\bibitem[Harris(1996)]{h96} Harris, W.E. 1996, \aj, 112, 1487
\bibitem[Moro \& Munari(2000)]{mm00} Moro, D. \& Munari, U. 2000, \aaps, 147, 361
\bibitem[Renzini \& Buzzoni(1986)]{rb86} Renzini, A. \& Buzzoni, A. 1986 in Spectral evolution of galaxies, eds.\ C.\ Chiosi \& A.\ Renzini (Dordrecht: Reidel) p.\ 195
\bibitem[Rodriguez-Merino et al.(2005)]{rcbb05} Rodriguez-Merino, L.H., Chavez, M., Bertone, E. \& Buzzoni, A. 2005, \apj, 626, 411
\bibitem[Santos \& Piatti(2004)]{sp04} Santos, J.F.C. Jr. \& Piatti, A.E. 2004, \aap, 428, 79
\bibitem[Thilker et al.(2005)]{th05} Thilker, D.A., Hoopes, C.G., Bianchi, L., et al. 2005, \apj, 619, L67
\bibitem[Tonry(1991)]{t91} Tonry, J.L. 1991, \apj, 373, L1
\bibitem[Tonry \& Schneider(1988)]{ts88} Tonry, J.L., \& Schneider, D.P. 1988, \aj, 96, 807
\bibitem[Worthey et al.(1994)]{w94} Worthey, G., Faber, S. M., Gonzalez, J.J. \& Burstein, D. 1994, \apjs, 94, 687
\bibitem[Wynne(1999)]{w99} Wynne, C.G. 1999, \mnras, 302, 830
\end{thebibliography}
\end{document}